\documentclass{llncs}
\usepackage{algorithm}
\usepackage{algpseudocode}
\usepackage{graphicx}
\usepackage{url}
\usepackage{hyperref}
\usepackage{cite}
\usepackage{array}
\usepackage{enumitem}
\usepackage{multicol}
\usepackage{booktabs}
\usepackage[english]{babel}
\usepackage{amssymb} 
\usepackage{mathtools}

\begin{document}    
    \title{Trustee: Full Privacy Preserving Vickrey Auction on top of Ethereum}
        \author{Hisham S. Galal and Amr M. Youssef}
        \institute{Concordia Institute for Information Systems Engineering,\\ Concordia University, Montr\'{e}al, Qu\'{e}bec, Canada}
    \maketitle{}
    \setcounter{page}{1}
    \pagenumbering{arabic}
    \pagestyle{plain}
    \begin{abstract}
    The wide deployment of tokens for digital assets on top of Ethereum  implies the need for powerful trading platforms. Vickrey auctions have been known to determine the real market price of items as bidders are motivated to submit their own monetary valuations without leaking their information to the competitors. Recent constructions have utilized various cryptographic protocols such as ZKP and MPC, however, these approaches either are partially privacy-preserving or require complex computations with several rounds. In this paper, we overcome these limits by presenting Trustee as a Vickrey auction on Ethereum which fully preserves bids' privacy at relatively much lower fees. Trustee consists of three components: a front-end smart contract deployed on Ethereum, an Intel SGX enclave, and a relay to redirect messages between them. Initially, the enclave generates an Ethereum account and ECDH key-pair. Subsequently, the relay publishes the account's address and ECDH public key on the smart contract. As a prerequisite, bidders are encouraged to verify the authenticity and security of Trustee by using the SGX remote attestation service.  To participate in the auction, bidders utilize the ECDH public key to encrypt their bids and submit them to the smart contract. Once the bidding interval is closed, the relay retrieves the encrypted bids and feeds them to the enclave that autonomously generates a signed transaction indicating the auction winner. Finally, the relay submits the transaction to the smart contract which verifies the transaction's authenticity and the parameters' consistency before accepting the claimed auction winner. As part of our contributions, we have made a prototype for Trustee available on Github\footnote{\url{https://github.com/hsg88/Trustee}} for the community to review and inspect it. Additionally, we analyze the security features of Trustee and report on the transactions' gas cost incurred on Trustee smart contract.
    \end{abstract}

\section{Introduction}
\par The wide success of Ethereum \cite{wood2014ethereum} with a market capitalization around 10 billion USD at the time of the writing \cite{coinmarket} has led to the deployment of thousands of asset-specific tokens \cite{tokenmarket}. Such a large-volume market demands powerful trading platforms. Auctions have been known to be an effective and efficient way to trade highly-valuable goods. Additionally, sealed-bid auctions have an important advantage compared to their open-cry counterparts. Precisely, given an honest auctioneer, bidders are assured that their competitors will not gain any information about their bids. Moreover, in a Vickrey auction which is a particular type of sealed-bid auctions, the auction winner pays the second highest-price. Consequently, Vickrey auctions motivate bidders to submit bids based on their own monetary valuation which essentially helps in determining the real market price of the auctioned items. Nonetheless, a corrupt auctioneer can easily compromise the aforementioned advantages. For instance, the auctioneer can (i) expose the bids' information to a colluding bidder, (ii) declare a false auction winner, (iii) set a fake second-highest price that is slightly lower than the highest price in order to gain an advantage.  Consequently, the major challenges in constructing a Vickrey auction are maintaining bids' privacy and verifying the correctness of the auction winner and the amount of the second-highest price.

\par Building a Vickrey auction on top of Ethereum to trade the deployed tokens essentially involves writing a \textit{smart contract} that adheres to a predefined protocol. A smart contract is an autonomous agent that resides at a specific address in the Ethereum blockchain. It contains functions to make decisions, and persistent storage to save state. The execution model of a smart contract is to lie passive and dormant until it is poked. More specifically, a smart contract only becomes active once any of its designated functions is invoked due to the receipt of either a \textit{message} from another smart contract, or a \textit{transaction} from an externally-owned account (i.e., informally called a \textit{wallet}). The lifetime of a smart contract is to exist as long as the whole Ethereum network exists unless it was programmed to \textit{self destruct} which essentially renders it completely inactive. With the help of the consensus protocol in Ethereum, a smart contract gains control flow integrity. In other words, it executes as its code dictates to the extent that even its creator cannot modify or patch it. The consensus protocol requires miners to do an expensive operation (\textit{proof of work}) in addition to processing and validating the transactions. Therefore, miners are compensated by a block reward in addition to transaction fees. Essentially, the more complex the transaction, the higher fees are incurred.  Additionally, processing and validating transactions imply that miners have a fully transparent access to smart contract's state. Therefore, the lack of privacy in addition to the expensive transaction fees are the main challenging issues in building a secure and efficient Vickrey auction on top of Ethereum.

\par To address the above issues, various constructions for sealed-bid auctions in general utilize different cryptographic protocols such as Zero-Knowledge Proofs (ZKP) and secure Multi-Party Computations (MPC) to ensure the verifiability of the auction winner without sacrificing bids' privacy. However, in the former, the auctioneer is an entity that learns bids' values and proves the correctness of the auction winner to the bidders. This approach is partial privacy-preserving since the bids' values are exposed to the auctioneer who may maliciously exploit this information in future auctions. In addition to the inherent high transaction fees in Ethereum, the verification of the auctioneer's proof is executed inside a smart contract which significantly incurs a high cost (e.g., zkSNARK verification roughly takes 3 million gas \cite{galal2018succinctly}) that renders the whole approach to be an expensive option. In contrast, the MPC approach can offer full bids' privacy at the cost of higher transactions fees since it requires several of complex computations between the bidders and using a smart contract as a public bulletin board in addition to an escrow of funds.
	 
\par We present Trustee as a trusted and efficient Vickrey auction on top of Ethereum that substantially overcomes the limitations of ZKP and MPC approaches. Trustee utilizes Intel Software Guard Extensions (SGX) \cite{anati2013innovative} as a Trusted Execution Environment (TEE) to fully preserve bids' privacy at a significantly cheaper transaction fee to verify the auction winner correctness compared to the aforementioned approaches. Intel SGX is a hardware architecture that provides an isolated  and tamper-proof environment called \textit{enclave}. In essence, the control flow integrity of the code and the confidentiality of the data inside an enclave are well protected from the host operating system and other running processes. Therefore, Intel SGX technology can complement smart contracts with confidential data processing, a highly desirable property that Ethereum lacks. 
\par Similar to other TEE technologies, Intel SGX has a poor availability, and its operation can be easily terminated at any point of time. Hence, a stateful application utilizing Intel SGX requires a storage with high availability such as the blockchain or IPFS \cite{benet2014ipfs} to persist sensitive state (e.g., the sealed-bids and sealed private keys).  We are also aware that several side-channel attacks on Intel SGX have been reported recently to leak information about the sensitive data inside enclaves such as private keys (more details in Section \ref{sec:design}). Therefore, we do strongly note that rather than building Trustee using only Intel SGX, we utilize a smart contract on Ethereum  for two purposes. First, it acts as an escrow to hold the initial deposits of bidders during the bidding phase for a specific time interval. As a result, bidders are not exposed to the theft of funds in the case they were sending their payments to an account controlled by the enclave which might get compromised. Secondly, it acts as a trusted judge that verifies on-behalf of the bidders the consistency of the inputs used by the enclave to determine the auction winner. Hence, it allows bidders with low-processing mobile devices to easily join the auction. Consequently, by integrating a smart contract on Ethereum with Intel SGX technology, Trustee becomes a robust Vickrey auction solution that inherits the best properties from the two worlds of blockchain and TEEs.
     \par \textbf{Our contribution}, we present the design and implementation of Trustee that provides the following properties:
    \begin{enumerate}
        \item {\textbf{Full privacy preserving}. The only information about bids that any bidder can learn besides to their own is the winning bid.}
        \item{\textbf{Cheap correctness verification cost}. Compared to other alternatives, Trustee achieves significantly cheaper verification cost of the auction winner correctness.}
        \item{\textbf{Rational fairness}. Malicious participants gain no advantage over honest parties. In fact, they are  obligated to follow the proposed protocol to avoid being financially penalized.}
        \item {\textbf{Efficiency}. The core computations of sealing bids, decrypting them, and selecting the auction winner are carried out in native environments off the blockchain which are more efficient than the Ethereum Virtual Machine (EVM).}
    \end{enumerate}
    We also provide an open-source prototype for Trustee  on Github (\url{https://github.com/hsg88/Trustee}) for the community to review it. The rest of this paper is organized as follows. Section \ref{sec:related}  provides a review of current constructions of sealed-bid auctions on top of blockchains and the integration of TEEs with blockchain. In Section \ref{sec:preliminaries}, we present the cryptographic primitives utilized in Trustee's design. Then, in Section \ref{sec:design}, we provide the protocol design behind Trustee, analyze its security features, and report the gas cost of the relevant transactions. Finally, we present our conclusions in Section \ref{sec:conclusion}.
    
    \section{Related Work}
    \label{sec:related}
    In this section, we provide a review of state-of-the-art constructions that utilize a variety of cryptographic protocols such as ZKP and MPC to build sealed-bid auction on top of blockchains. Then, we briefly present recent works that integrate blockchain with TEE to provide elegant solutions.
    \subsection{Sealed-Bid Auctions On Blockchain}
    \par Blass and Kerschbaum \cite{blass2018strain} proposed \textit{Strain} as a protocol to build a sealed-bid auction on top of the blockchain technology. Strain utilizes a two-party computation protocol to compare pairs of bids, and the outcome is stored on a blockchain. Additionally, Strain utilizes ZKP to prove that the outcome is correct with respect to the compared pairs of bids. Strain fully preserve bids' privacy. However, its complexity scales proportionally to the number of bidders. Moreover, as reported by its authors, it reveals the order of the bids as it behaves similar to Order-Preserving Encryption (OPE) schemes.
    
    \par Galal and Youssef \cite{galal2018verifiable} proposed a protocol that utilizes Pedersen commitment and Honest-Verifier Zero-Knowledge (HVZK) range proof to build a public verifiable sealed-bid auction on top of Ethereum.  During the bidding phase, the bidders submit Pedersen commitments of their bids to the auction smart contract. Then at the reveal phase, they open their commitments individually to the auctioneer using RSA public-key encryption. Finally, the auctioneer declares the auction winner and utilizes HVZK range proof with the auction smart contract as a verifier to prove the correctness of the auction winner. However, the protocol has the following issues: (i) running an interactive HVZK with a smart contract as a verifier is not secure due to the possible influence of miners on the challenge step, (ii) the proof size and verification cost scales proportionally with the number of bidders, and (iii) the protocol is partial privacy-preserving as the auctioneer gains knowledge of all bids values.
    
    \par Motivated to improve on their latest work, Galal and Youssef \cite{galal2018succinctly} utilized Zero-Knowledge Succinct Non-interactive Argument of Knowledge (zkSNARK) \cite{ben2014succinct} which is an innovative cryptographic method in the field of Verifiable Computation. In contrast to their previous work \cite{galal2018verifiable}, this protocol has several desirable properties that synergies with the blockchain technology: (i) a constant short-size proof, (ii) a constant verification cost, (iii) a non-interactive protocol that takes one message to convince the verifier (i.e., the smart contract).  However, generating a zkSNARK proof scales proportionally with the number of multiplication gates in the arithmetic circuit of their computation problem which further depends on the number of bidders. Moreover, the protocol assumes a trusted setup of the proving and verification keys. Finally, the protocol is a partial privacy-preserving  where bidders have to trust the auctioneer to not exploit their bids values in future auctions.
    
    \subsection{SGX with Blockchain Solutions}
    \par Several recent constructions utilized TEE technologies such as Intel SGX to solve privacy and performance issues on the blockchain, (e.g., see \cite{zhang2016town, cheng2018ekiden, tran2017obscuro, lind2016teechan, bentov2017tesseract, al2018airtnt, milutinovic2016proof}). In here, we provide a brief review of the works that Trustee shares some similarities with. In  \cite{zhang2016town} Zhang et al. proposed Town Crier (TC): an authenticated data feed that gives smart contracts on Ethereum the ability to request data from existing HTTPS-enabled data sources. TC consists of three components: a front-end smart contract, a back-end Intel SGX enclave, and a relay to redirect messages between them. Initially, the TC's front-end receives a request from a smart contract on Ethereum. The relay monitors the Ethereum blockchain for such a request and forwards it to TC's back-end.  Then, the TC's back-end resolves this request and outputs a transaction containing the response. Finally, the relay submits the transaction to TC's front-end where it triggers the execution of a callback on the relying smart contract.
    
    \par Cheng et al. \cite{cheng2018ekiden}  proposed Ekiden: a platform for confidentiality-preserving, trustworthy, and performant smart contract execution to solve the inherent lack of privacy and poor performance in blockchains. Ekiden's architecture separates smart contract execution from the consensus protocol. It preserves the confidentiality of a smart contract's states, besides to, achieving high throughput and scalability. The authors evaluated a prototype (with Tendermint as the consensus layer) and reported a performance of 600x more throughput and 400x less latency at 1000x less cost than the Ethereum mainnet. 
    
    \par Tran et al. \cite{tran2017obscuro} proposed Obscuro: an Intel SGX-backed mixer to address the anonymity issue on Bitcoin. Due to the pseudo-anonymity offered by Bitcoin, the link between the transaction's sender and receiver can be exploited to cluster and track users which defeats the goal of anonymous payment. Obscuro utilizes Intel SGX to preserve the privacy of the mixer's participants and perform a secure shuffle of bitcoins. Users post their deposits indirectly on Bitcoin blockchain rather than directly interacting with Obscuro. Consequently, malicious operators cannot prevent benign users from mixing their bitcoins. Furthermore, Obscuro does not store any operation states outside of the TEE to counter the possibility of state-rewind in conjunction with eclipse attacks. The authors evaluated Obscuro on Bitcoin testnet and reported that they were able to mix 1000 inputs in just 6.49 seconds.

    \section{Preliminaries}
    \label{sec:preliminaries}
    In this section, we briefly introduce the cryptographic primitives that are utilized in our design for Trustee.
    
    \par Ethereum utilizes Elliptic Curve Digital Signature Algorithm (ECDSA) to verify the authenticity of transactions. To create an account on Ethereum, one has to statistically randomly generate a unique ECDSA key-pair $(pk, sk)$ on the curve \textit{secp256k1} \cite{secp256k1,bos2014elliptic}. Keeping the private key secure is essential because it is used to sign transactions originating from the associated account. The address of an account is  the rightmost 20-bytes of the \textit{Keccak256} \cite{bertoni2013keccak} hash of the public key. This results in a more compact address size compared to the 64-bytes public key.  When a transaction is sent to the network, miners are tasked with verifying the transaction's signature with respect to the sender's address. Precisely, ECDSA consists of the following three algorithms:
    \begin{enumerate}
        \item {$(pk, sk) \leftarrow  \texttt{Gen($1^\lambda$)}$ which generates the public key  $pk$ and the associated  private key $sk$ based on the security parameter $\lambda$.}
        \item {$\sigma \leftarrow \texttt{Sign(H(m),sk)}$ which generates the signature $\sigma$ for the hash of the message $m$ under a designated hash function $H$ and the private key $sk$.}
        \item{ $(\top / \bot) \leftarrow \texttt{Verify}(\sigma,H(m), pk)$ which verifies the signature $\sigma$ on the hash of message $m$ under the public key $pk$.}
    \end{enumerate}

    \par The second cryptographic protocol we utilize is Elliptic Curve Integrated Encryption Scheme (ECIES) \cite{gayoso2010survey}. It enables two parties to communicate  authenticated confidential messages. As its name indicates, ECIES integrates the following functions: 
    \begin{enumerate}
    \item{$(sk,pk) \gets \texttt{KGen}(params)$: a key generation function that takes elliptic curve parameters $params$ to produce a random private key $sk$ and the associated public key $pk$.}
        \item {$ss \gets \texttt{KA}(sk_i, pk_j)$: a key agreement function to generate a shared secret $ss$ based on the private key of party $i$ and the public key of party $j$.}
        \item {$(k_1, k_2) \gets \texttt{KDF}(ss)$: a key derivation function  to produce keys $k_1$ and $k_2$ from the shared secret $ss$.}
        \item {$ct \gets \texttt{Enc}_{k_1}(m)$: a symmetric encryption function to encrypt a message $m$ using the symmetric key $k_1$.}
        \item {$tag \gets \texttt{MAC}_{k_2}(m)$: a message authentication code function to generate a tag based on the key $k_2$ and the message $m$.}
    \end{enumerate}
    To demonstrate how ECIES works, assume that Alice wants to encrypt a message $m$ and send it to Bob. They initially agree on common ECIES parameters $params$. Then, Alice and Bob  individually generate the ephemeral key pairs $(sk_A, pk_A), (sk_B, pk_B)$, respectively. Subsequently, Alice does the following steps:
     \begin{enumerate}
     \item{Create a shared secret $ss \gets \texttt{KA}(sk_A, pk_B)$}
     \item{Derive two keys $(k_1, k_2) \gets \texttt{KDF}(ss)$.}
     \item{Obtain the ciphertext of her message $ct \gets \texttt{Enc}_{k_1}(m)$.}
     \item{Authenticate the ciphertext by creating a $tag \gets \texttt{MAC}_{k_2}(ct)$.}
     \item{Send the tuple $(pk_A, ct, tag)$ to Bob.}
     \end{enumerate}
     Once Bob receives the tuple $(pk_A, ct, tag)$, he can decrypt the ciphertext and verify its authenticity by doing the following:
     \begin{enumerate}
         \item {Create a shared secret $ss \gets \texttt{KA}(sk_B, pk_A)$}
     \item{Derive two keys $(k_1, k_2) \gets \texttt{KDF}(ss)$.}
     \item{Assert that $tag = \texttt{MAC}_{k_2}(ct)$, otherwise, he rejects.}
     \item{Obtain the message $m \gets \texttt{Enc}^{-1}_{k_1}(ct, k_1)$.}
     \end{enumerate}

    \section{Trustee's Design and Analysis}
    \label{sec:design}
    In this section, we briefly present the architecture of Trustee and illustrate the interaction flow between its components. Then, we explain the protocol in details. Next, we mention the threat model, security assumptions, and elaborate by analyzing various possible adversary attacks. Finally, we provide the implementation details of Trustee's prototype and evaluate the transactions gas costs. 
    \subsection{Trustee's Architecture}
\begin{figure}
    	\includegraphics[scale=.53]{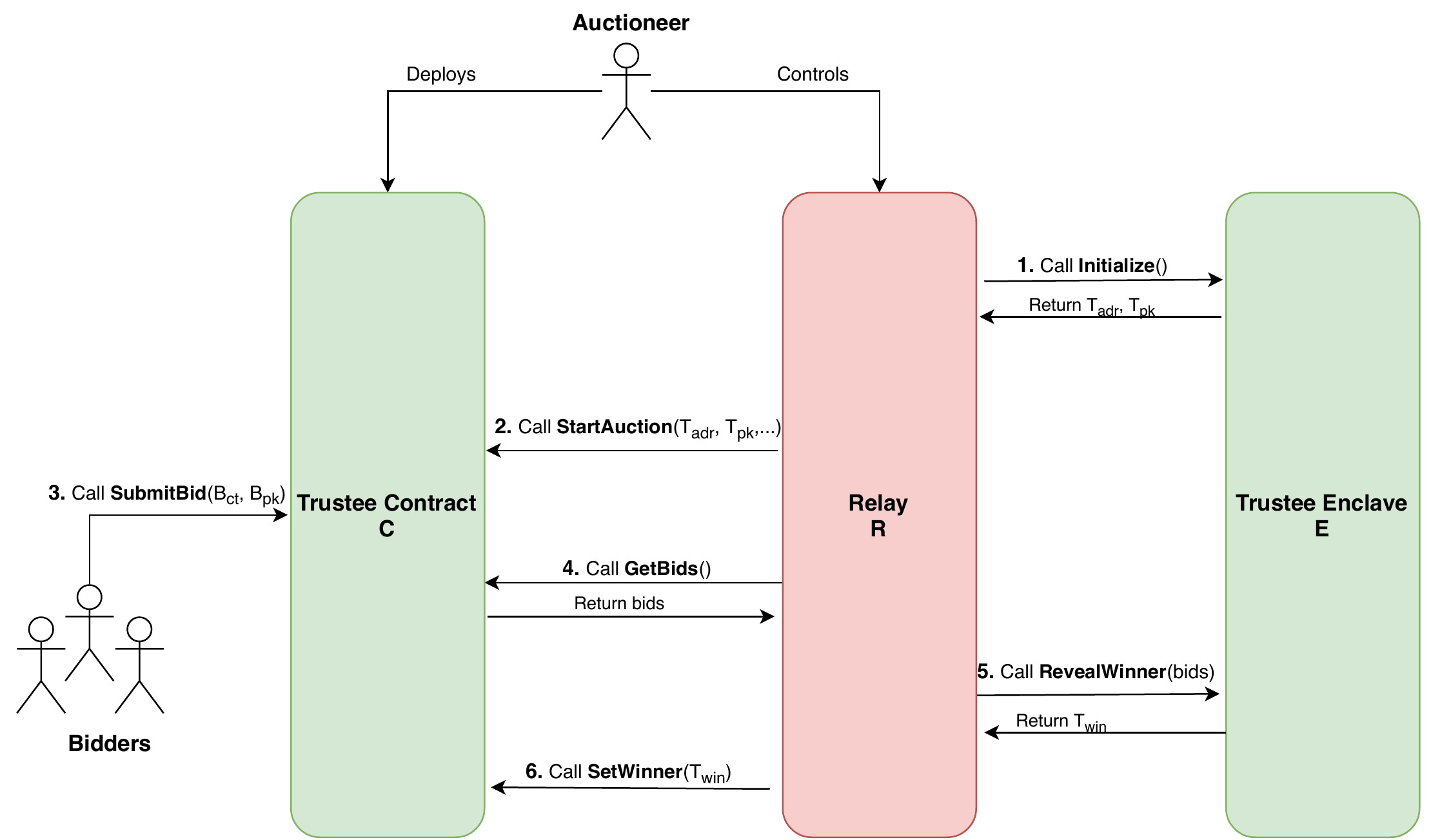}
    	\caption{Interactions between Trustee's components and bidders. The green components are trusted}
    	\label{figArch}
    \end{figure}

    Trustee consists of three components: a smart contract $C$ which resides on top of  Ethereum, a back-end Intel SGX enclave $E$ and a relay $R$ which both run off-chain on a server. We refer to the user who deploys $C$ and controls $R$ as the auctioneer. Furthermore, $E$ is only accessible through $R$, and $R$ interacts with $C$ on behalf of the auctioneer and $E$. The general flow of interactions between Trustee's components, and bidders is depicted in Fig. \ref{figArch}. 

    \par Initially, the auctioneer deploys $C$ on Ethereum and publishes its address so that interested sellers and buyers can learn about it. To start an auction, the auctioneer sends a request to $R$ which loads $E$ and calls the function \texttt{Initialize()}. As a response, $E$ generates an externally owned Ethereum account  with the private key $T_{sk}$ and the associated address $T_{adr}$, and an ECDH key-pair $(T_{dh}, T_{pk})$ where $T_{dh}$ is the private key and $T_{pk}$ is the associated public key. Then, it returns the values of $T_{adr}$ and $T_{pk}$ to $R$. Subsequently, the auctioneer instructs $R$ to set the stage for a new auction on $C$ by calling the function \texttt{StartAuction} which takes $T_{adr}$ and $T_{pk}$. Next, assume a bidder Bob is interested in the auction, then he utilizes ECIES protocol with $T_{pk}$ as the public key of the recipient (i.e., Trustee's enclave $E$) to seal his bid. Subsequently, he submits his sealed bid $B_{ct}$ along with his ECDH public key $B_{pk}$ to $C$. Once the bidding interval is closed, $R$ retrieves the sealed bids stored on the $C$, then it forwards them to $E$ by calling the function \texttt{RevealWinner}. As a result, $E$ opens the sealed bids and determines the winner and second-highest price. Then, it returns a transaction $T_{win}$ signed by the private key $T_{sk}$ to $R$. Finally, $R$ sends $T_{win}$ to $C$ which is essentially a call to the smart contract function $SetWinner$ that declares the auction winner and second-highest price. 
    
    \subsubsection{Initializing an Auction}
     The initialization process starts with the auctioneer requesting $R$ to load $E$ inside Intel SGX enclave and invoke the function \texttt{Initialize()} which is implemented as shown in Algorithm: \ref{alg:boot}.
      \begin{algorithm}
    	\caption{Initializing State of Trustee's Enclave}
    	\label{alg:boot}
    	\begin{algorithmic}[1]
    		\Function{Initialize}{}
    		\State $(T_{pk}, T_{\text{dh}}) \leftarrow $\texttt{GenerateECDHKeys()}
    		\State $(T_{adr}, T_{sk}) \leftarrow $\texttt{GenerateAccount()}
    		\State $sealedState \leftarrow \texttt{Seal}(T_{sk}, T_{dh})$
    		\State $\texttt{return }(sealedState, T_{adr}, T_{pk})$
    		\EndFunction
    	\end{algorithmic}
    \end{algorithm} 
\par The \texttt{Initialize()} function generates two key-pairs. More precisely, one key-pair $(T_{pk}, T_{dh})$ that enables bidders to seal their bids such that only $E$ can open them, and the second one to authenticate the result (i.e., auction winner and second-highest price) generated by $E$. The former is an ECDH key-pair used as part of ECIES protocol between $E$ and each bidder to securely transmit the sealed bids through $C$ and $R$. The later is an ECDSA key-pair used to sign the result. Verifying the signature on the result by $C$ is a relatively expensive operation (i.e., roughly 120,000 gas for using \textit{ecrecover}). Therefore, in Trustee, we utilize an intrinsic operation that happens on every transaction in Ethereum (i.e., transaction's signature verification) to indirectly verify the authenticity of the result for us. Hence, $E$ generates an ECDSA key-pair on curve \texttt{secp256k1} which essentially creates  an external owned Ethereum account with the private key $T_{sk}$ and the associated address $T_{adr}$. Then, whenever $E$ determines the auction winner and the second-highest price, it outputs a transaction $T_{win}$ signed by $T_{sk}$. Later, $R$ sends $T_{win}$ to the Ethereum network, where the miners verify its signature. Finally, $C$ only has to assert that the sender of $T_{win}$ is the $T_{adr}$. As a result, this approach yields a  much cheaper transaction fee  compared to the explicit signature verification by calling \texttt{ecrecover}.
   
\par Intel SGX enclaves are designed to be stateless. In other words, once an enclave is destroyed, its whole state is lost. However, in Trustee, we have to persist the generated keys as long as the current auction is running. Therefore, we utilize Intel SGX feature known as \textit{Sealing} \cite{anati2013innovative} to properly save the generated private keys.  Sealing is the process of encrypting enclave secrets in order to persist them on a permanent storage such as a disk. This effectively allows us to retrieve the private keys $(T_{sk}, T_{dh})$  even if the enclave was brought down for any reason. The encryption is performed using a private \textit{Seal Key} that is unique to the  platform and enclave, and is not accessible by any other entity. 

\par Upon the return from \texttt{Initialize}, $R$ saves the values of $sealedState$ on a disk besides to having a backup. Furthermore,  $R$  publishes the values $T_{adr}$ and $T_{pk}$ by calling the function \texttt{StartAuction} on $C$ as shown in Fig. 2. The function \texttt{StartAuction} also takes extra parameters that control the different intervals of the current auctions. More precisely, $T_1, T_2$, and $T_3$ which define the numbers of the blocks before which: (i) bidders submit their sealed bids, (ii) $R$  submits $T_{win}$, (iii) honest participants (i.e., auctioneer and non-winning bidders) reclaim the initially deposited fund $D$, respectively. The initial deposit $D$ is paid by all participants to penalize malicious behavior. 

\subsubsection{Provisioning of Bids}
Once the new auction has been initialized, an interested bidder Bob can seal his bid $x$  by utilizing ECIES as shown in Algorithm \ref{alg:ecies}. It starts with retrieving the public key $T_{pk}$ from $C$. Then, it generates an ephemeral ECDH key-pair $(B_{pk}, B_{sk})$ on \texttt{curve25519} where $B_{sk}$ is the private key and $B_{pk}$ is the associated public key. Then, it computes the shared secret $s$ based on $T_{pk}$ and $B_{sk}$. After that, it derives two symmetric keys $k_1$ and $k_2$ in order to perform an authenticated encryption on the bid value $x$. Finally, it returns the sealed-bid $B_{ct}$ and the associated public key $B_{pk}$. Subsequently, Bob sends the values $B_{ct}$ and $B_{pk}$ to the function \texttt{SubmitBid} on $C$ as shown in Fig. 2.
\begin{center}
		\fbox{\centering
			\begin{tabular}{p{.2\linewidth}p{.75\linewidth}}
				\textbf{StartAuction:}& upon receiving $(T_{adr}, T_{pk}, T_1, T_2, T_3, D)$ from auctioneer \textbf{A}  \\ 
				&       \texttt{Assert} $state = Init$\\
				&       \texttt{Assert} $ledger[A] >= D$\\
				&       \texttt{Set} $ledger[A] := ledger[A] - D$\\
				&       \texttt{Set} $deposit := deposit + D$\\
				&       \texttt{Set} $state := Bidding$\\
				&       \texttt{Store} $T_{adr}, T_{pk}$\\
				&       \texttt{Store} $T_1, T_2, T_3, D$\\
				\textbf{SubmitBid:}& upon receiving $(B_{ct}, B_{pk})$ from a bidder  \textbf{B}  \\ 
				&       \texttt{Assert} $state = Bidding$\\
				&       \texttt{Assert} $T < T_1$\\
				&       \texttt{Assert} $ledger[B] >= D$\\
				&       \texttt{Set} $ledger[B] := ledger[B] - D$\\
				&       \texttt{Set} $ledger[C] := ledger[C] + D$\\
				&       \texttt{Set} $bids[B] := (B_{ct}, B_{pk})$\\
				&       \texttt{Set} $Bidders := Bidders \cup \{B\}$\\
				\textbf{SetWinner:}& upon receiving $(H, I, P)$ from the an address  \textbf{X}  \\ 
				&       \texttt{Assert} $X = T_{adr}$\\
				&       \texttt{Assert} $state = Bidding$\\
				&       \texttt{Assert} $T_1 < T < T_2$\\
				&       \texttt{IF Keccak256}$(bids.B_{ct} || bids.B_{pk}) \neq H$\\
				&       \hspace{.75cm}\texttt{Set} $state := Rejected$\\
				&       \hspace{.75cm}\texttt{Return}\\
				&       \texttt{EndIF}\\
				&       \texttt{Set} $state := Revealed$\\
				&       \texttt{Set $winner = Bidders[I]$}\\
				&       \texttt{Set $price = P$}\\
				\textbf{Withdraw:}& upon receiving $()$ from an address   \textbf{X}  \\ 
				&       \texttt{Assert} $T_2 < T < T_3$\\
				&       \texttt{IF} $(state = Revealed \textbf{ and } X \in \{A\} \cup \{Bidder\} - \{winner\})$\\
				&       \texttt{    }$\textbf{ or } (state = Rejected \textbf{ and } X \in \{Bidder\})$\\
				&       \hspace{.75cm}\texttt{Set }$ledger[C] := ledger[C] - D$\\
				&       \hspace{.75cm}\texttt{Set }$ledger[X] := ledger[X] + D$\\
				&       \texttt{EndIF}\\
				\textbf{Reset:}& upon receiving $()$ from the auctioneer \textbf{A}\\
				&       \texttt{Assert} $T_3 < T$\\
				&       \texttt{Set} $state := Init$\\
				&       \texttt{Clear} $Bidders$\\
				&       \texttt{Clear} $bids$\\
		\end{tabular}} 
	\end{center}
	\par\textbf{Figure 2.} Pseudocode for the Trustee's smart contract $C$\\

 \begin{algorithm}
    	\caption{Sealing of Bids using ECIES}
    	\label{alg:ecies}
    	\begin{algorithmic}[1]
    		\Function{SealBid}{x}
    		\State $T_{pk} \gets $\texttt{GetTrusteePublicKey}$()$
    		\State $(B_{pk}, B_{sk}) \gets $ \texttt{GenerateECDHKeys}$()$
    		\State $s \gets $\texttt{ComputeSharedSecret}$(KA(B_{sk}, T_{pk})$
    		\State $(k_1, k_2) \gets $ \texttt{DeriveKeys}$(s)$
    		\State $iv \gets $ \texttt{InitRandomIV}$()$
    		\State $ct \gets $ \texttt{Encrypt}$(x, iv, K_1)$
    		\State $tag \gets $ \texttt{MAC}$(ct, K_2)$
    		\State $B_{ct} \leftarrow ct|| iv || tag$
    		\State \texttt{return }$ (B_{ct}, B_{pk})$
    		\EndFunction
    	\end{algorithmic}
    \end{algorithm} 

    \par The function \texttt{SubmitBid} first asserts that: (i) the current state is set to \texttt{Bidding}, and (ii) the call is invoked before the end of the bidding interval. After that, it deducts the initial deposit $D$ from Bob and stores the $B_{ct}$ and $B_{pk}$ into the array $bidders$. Note that the size of the $B_{ct}$is 32 bytes. Moreover, we utilize the Curve25519 for generating ECDH key-pairs  due to two reasons: (i) it only uses compressed elliptic point (i.e., $X$ coordinate), so it provides fast and efficient ECDH, (ii) the public-key size becomes 32-bytes rather than 64-bytes, therefore, both the $B_{ct}$ and $B_{pk}$ synergies effectively with Ethereum native variable type $uint256$.
    
    \subsubsection{Revelation of the Auction Winner }
    Once the bidding interval is over, $R$ retrieves the submitted array of sealed bids $B_{ct}$ and their associate public keys $B_{pk}$ from $C$. Then, it passes them along with ${sealedState}$ (previously generated by the function \texttt{Initalize}) to the function \texttt{RevealWinner} on $E$ as shown in  Algorithm \ref{alg:reveal}. 
    \begin{algorithm}
    	\caption{Revelation of the Auction Winner}
    	\label{alg:reveal}
    	\begin{algorithmic}[1]
    		\Function{RevealWinner}{$B_{ct}[], B_{pk}[], sealedState$}
     		\State $max \gets 0$
     		\State $second \gets 0$
    		\State $index \gets -1$
    		\State $N \gets $\texttt{Length}$(B_{ct})$
    		\State $(T_{sk}, T_{dh}, success)= \texttt{Unseal}(sealedState)$
    		\If{ $success = 0$}
    		\State \Return{}
    		\EndIf
    		\For {$i \gets 1 \text{ to }  N$}
    		\State $bid \leftarrow \texttt{Decrypt}(B_{ct}[i], B_{pk}[i],  T_{dh})$
    		\If {$max < bid$}
    		\State $second \gets max$
    		\State $max \gets bid$
    		\State $index \gets i$
    		\EndIf
    		\EndFor
    		\State $hash \gets \texttt{Keccak256}(B_{ct} || B_{pk})$
    		\State {$T_{win} \gets $\texttt{CreateTransaction}$(C, hash, index, second, T_{sk})$}
    		\State {$sealedState \gets \texttt{Reset()}$}
    		\State \Return {$(T_{win}, sealedState)$}
    		\EndFunction
    	\end{algorithmic}
    \end{algorithm} 
    In this function, $E$ initially unseals the private keys from $sealedState$. Then, for every bidder $i$, it runs the decryption part of ECIES protocol based on  the sealed-bid $B_{ct}[i]$, the  public key $B_{pk}[i]$, and  the private key $T_{dh}$  to extract the bid value and find the winner. Once all sealed-bids $B_{ct}$ are decrypted, the winner's index and second-highest bid are set accordingly in the variables $index$ and $second$. Subsequently, $E$ binds the auction winner to the inputs it received by computing the \texttt{Keccak256} hash value of  $B_{ct}$ and $B_{pk}$. Finally, $E$ creates a raw transaction $T_{win}$ with the destination address as $C$ and signs it with the private key $T_{sk}$. For the sake of simplicity, we defer explaining the details of \texttt{Reset()} and \texttt{Unseal()} to Subsection \ref{sec:analysis}.
    
    \par Subsequently, the auctioneer has to send some funds to $T_{adr}$ in order to pay the transaction fees to be incurred by $T_{win}$. Next, the auctioneer requests $R$ to send the transaction $T_{win}$ to $C$ which is essentially a call to the function \texttt{SetWinner} shown in the Fig. 2. It takes the following parameters: (i) $H$ as \texttt{Keccak256} hash value of the inputs $B_{ct}$ and $B_{pk}$, (ii) as the index of the winner in the array $Bidders$ which is further used by $C$ to determine the address of the auction winner, and (iii) $P$ as the second-highest price. On its call, it asserts that: (i) $T_{win}$'s origin is the address $T_{adr}$, the call happens within the auction winner revelation interval, and (iii) the $state$ is set to $Bidding$. Then, it checks if $H$ is equal to the \texttt{Keccak256} hash value of the  sealed bids and their associated public keys submitted by bidders. Accordingly, it decides whether to accept the submitted values or reject them. Eventually, it reflects the decision on its $state$. 
    
    \par Honest participants can reclaim their initial deposits within the withdraw interval by calling the function \texttt{Withdraw} shown in Fig. 2. Additionally, in the case of a successful winner revelation, then the winner's initial deposit is locked to set the stage for payment of the winning bid. Eventually, after the withdraw interval, the auctioneer calls the function $Reset$ in order to set the $state$ of $C$ to $Init$ so that new auctions can be started later by calling $StartAuction$.
    

    \subsection{Threat Model}
    \label{sec:analysis}
    In Trustee threat model, we assume the following:
    \begin{enumerate}
        \item {The smart contract $C$ is deployed on the \textit{mainnet} of Ethereum with an open-source code that is available for all bidders. Moreover, the functions on $C$ process the input parameters of transactions as their code dictate which is essentially enforced by Ethereum. Furthermore, all transactions in Ethereum are authenticated such that $C$ can precisely determine the sender address.}
        \item {The enclave $E$ is loaded inside a properly implemented and manufactured Intel SGX platform. Additionally, the source code of $E$ is available for all bidders. Finally, $E$ is properly programmed such that it does not have a bug that compromises the confidentiality of sealed-bids and private keys.}
        \item{The relay $R$ is the only interface to $E$ and is controllable by the auctioneer. The bidders have a black-box view of $R$ (i.e., closed-source code). Furthermore, $R$ is potentially untrusted component that can behave maliciously to compromise the security of Trustee.}
        \item{The Adversary is financially rational and powerful enough to have access to the host running $E$ and $R$. Hence, the adversary is able to control the execution of privileged software such as the operating system and the network-stack driver. However, the adversary cannot compromise the security model of Ethereum in order to maliciously change the state of $C$.}
    \end{enumerate}
    We acknowledge that several recent studies have uncovered side-channel attacks to compromise the confidentiality of Intel SGX \cite{vanbulck2018foreshadow, weisse2018foreshadowNG, xu2015controlled,schwarz2017malware, lee2017inferring,chen2018sgxpectre}. Also, multiple mitigation techniques have been proposed to address attack-specific issues \cite{shih2017t, shinde2016preventing, seo2017sgx, gruss2017strong}. Resolving side-channel attacks on Intel SGX enclave is beyond the  scope of this paper and is left for future work.
    \subsection{Security Analysis}
    We discuss the security of Trustee against possible scenarios including Intel SGX masquerade, eclipse, fork, and replay attacks \cite{brandenburger2018blockchain}. 
    \par
\begin{figure}
    	\includegraphics[scale=.55]{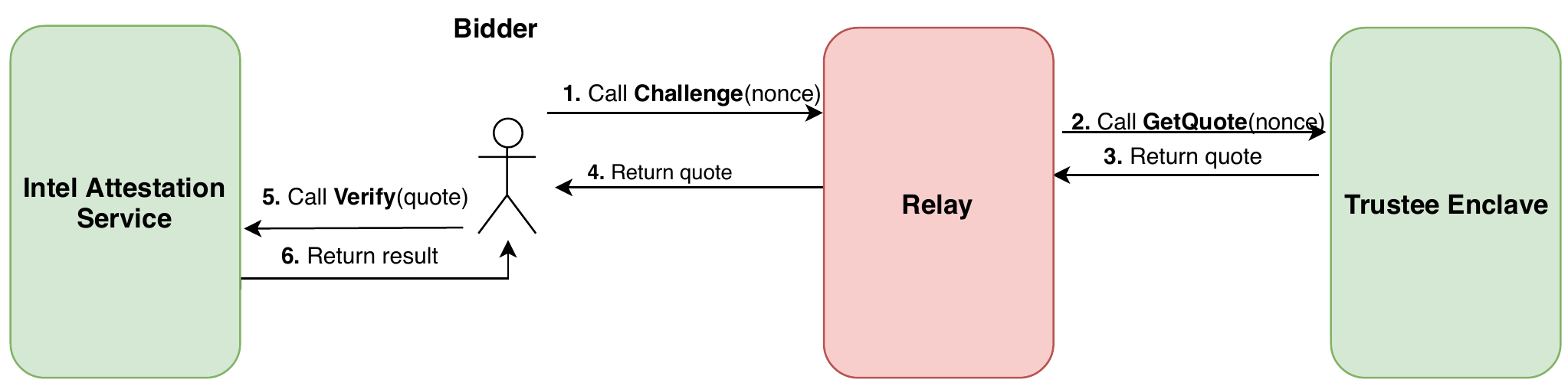}
    	\caption{Remote attestation of Trustee's Enclave}
    	\label{attestation}
    \end{figure}

 \noindent\textbf{Intel SGX Masquerade.} Since bidders do not have direct access to Trustee's enclave $E$, a corrupt auctioneer might generate the private keys and post the corresponding public key $T_{pk}$ and address $T_{dh}$ on the smart contract $C$. Incautious bidders would seal their bids by $T_{pk}$ which effectively gives the corrupt auctioneer access to the underlying bids. To counter this attack, we show how a wary bidder Bob can verify that the private keys $(T_{sk}, T_{dh})$ were generated by $E$  inside a genuine Intel SGX enclave. Essentially, Bob has to do the verification before submitting his sealed-bid. Therefore, once a new auction is started by the function \texttt{StartAuction}, Bob and Trustee engage in a protocol that utilizes the  \textit{Remote Attestation} \cite{anati2013innovative} feature of Intel SGX as shown in Fig. \ref{attestation}.
    Initially, Bob challenges $E$ through $R$ by calling the function \texttt{Challenge} and passes a $nonce$ to it. Then, $R$ forwards the $nonce$ to $E$ by calling the function \texttt{GetQuote}. Inside \texttt{GetQuote}, $E$ binds  $T_{adr}, T_{pk},$ and $nonce$ by hashing their concatenation and creating a digest $h \leftarrow \text{SHA256}(T_{adr}||T_{pk}||nonce)$. Then, it embeds $h$ as a user data  into a report $r$ by calling an Intel SGX supplied function \texttt{sgx\_create\_report}. Finally, $R$ passes $r$ to an Intel provided enclave known as the Quoting Enclave (QE) which verifies $r$ then signs it with Intel Enhanced Privacy ID (EPID) secret key to yield a \textit{quote}.  The Intel EPID  is device-specific and is only accessible by the QE. Subsequently, $R$ returns the quote to Bob who in turn contacts Intel Attestation Service (IAS) to verify the quote's signature. On a successful verification, Bob has to check the following: (i) the quote's user data is equivalent to $h$, and (ii) the source code of $E$ when compiled produces the same measurement (i.e., a digest of code and data of $E$) included in the quote.  Assuming IAS to behave honestly, then it is computationally infeasible for the adversary to generate a quote that asserts the authenticity of $E$ on a fake Intel SGX enclave to Bob.
   	\par \noindent\textbf{Eclipse Attack.} Generally, Intel SGX enclaves do not have trusted access to the network; therefore, Trustee's enclave $E$ is oblivious of the current state (i.e, sealed bids) on the smart contract $C$. Consequently, a corrupt auctioneer can provide an arbitrary subset of the sealed bids to $E$ in order to give advantage to a cartel of colluding bidders. A trivial solution to this challenge is to embed a full-node Ethereum client inside $E$ such that it can verify the PoW (Proof of Work) of Ethereum blocks and determine the correct state of the smart contract $C$. This solution is computationally secure against an adversary who controls less than $51\%$ of the hash rate power of the network. However, the TCB of $E$ becomes bloated with and susceptible to bugs founds in the client source code. Alternatively, in Trustee, we bind the output (i.e., winner's index and second-highest price) to the input (i.e., the set of  sealed-bids and associated public keys) by including the hash of the input as a parameter in the transaction $T_{win}$ as shown in Algorithm \ref{alg:reveal}. Therefore, the smart contract $C$ can determine whether all or a subset of the sealed bids were provided to $E$ by comparing hash parameter of $T_{win}$ to the hash of all bids and associated public keys in its state as shown in the function \texttt{SetWinner} in Fig. 2. 
    \par \noindent\textbf{Replay and Fork Attack.}
    We assess the possibility of a corrupt auctioneer Eve trying to compromise the privacy of the sealed-bids without being noticed and penalized. Recall that in the design of Trustee, $R$ initially calls the function \texttt{Initialize}, then at a later point in time, it calls the function \texttt{RevealWinner} to finalize the auction. The idea behind this attack is that Eve can launch multiple instances of $E$ and replay the same $sealedState$ to all instances but provide different subsets of the sealed-bids. Obviously, Eve gives one of the instances the correct number of sealed-bids  and its output is forwarded to $C$ to avoid penalty as discussed above. However, for the other instances she simply learns the outputs and discard them which effectively gives her access to all the underlying bids values.
    \par To counter this attack, we enforce Trustee's design of using fresh $sealedState$ for every call to the function \texttt{RevealWinner} by utilizing Intel SGX non-volatile hardware monotonic counters. Simply, the  function \texttt{Seal} called inside the function \texttt{Initialize} increments and reads the monotonic counter $ctr$, then it combines $ctr, T_{sk}$, and $T_{dh}$ and seals them into $sealedState$. Later, when the function \texttt{RevealWinner} is called, it invokes the function \texttt{Unseal} which unseals $sealedState$, then it reads the current monotonic counter and compares it with the unsealed $ctr$. Hence, if the equality check passes, then the function \texttt{RevealWinner} increments the counter as well and proceeds to the next steps, otherwise, it aborts without determining the auction winner (i.e., returning  an empty $T_{win}$ that does not indicate the auction winner.) Consequently, Eve can get valid output from \texttt{RevealWinner} only one time per a single auction regardless of how many instances of $E$ are launched. Alternatively, to avoid the low performance of using monotonic counters which takes approximately $80$ to $200$ ms for read/write operation, we can utilize a distributed system of Intel SGX enclaves to manage the state freshness as explained in \cite{matetic2017rote}.

    \subsection{Prototype Implementation and Gas Cost Analysis}
    Intel SGX cryptographic library does not support the curves \textit{secp256k1} and \textit{curve25519}, so we  utilize an Intel SGX compatible port of \textit{mebdtls} library \cite{mbedtls} as a static enclave library linked to Trustee's enclave. Mbedtls library is mainly used in ECDH and ECDSA key generation, ECDH shared secret derivation, and ECDSA signing. 
    We evaluate Trustee on a  Dell Inspiron 7577 laptop that is  SGX-enabled with the 6th Generation Intel Core i5 CPU and 8-GB of memory. We enable Intel SGX feature on the laptop's BIOS and allocate maximum allowed 128-megabytes memory for individual SGX enclave. 
    Also, we implement Trustee's smart contract in \textit{Solidity} which is the de-facto programming language for developing smart contracts in Ethereum.  Furthermore, we utilize \textit{Ganache} to set up a personal Ethereum blockchain in order to run tests, execute commands, and inspect state while controlling how the chain operates. 
    
    We report on the gas cost of transactions in Trustee for a Vickrey auction with $N = 100$ bidders and compares it with approaches in \cite{galal2018verifiable,galal2018succinctly} in Table \ref{tab.cost}. At the time of writing, December 14th, 2018, the median gas price is 3.3 \textit{GWei} and the average exchange rate for $1$ ether = \$83 USD. In other words, 1 million gas incurs transaction fees $\approx$ \$0.27 USD.
    \begin{table}[H]
    	\centering
    	\caption{\small{Gas cost of transactions in Trustee and auctions \cite{galal2018verifiable,galal2018succinctly} }}
    	\label{tab.cost}
    	\begin{tabular}{@{}l|r|r|r@{}}
    		\toprule
    		Function&Trustee& Auction \cite{galal2018verifiable}& Auction \cite{galal2018succinctly} \\
    		\midrule
    		\texttt{Deployment} &1173779&3131261&1346611 \\
    		\texttt{StartAuction} & 188201&---&--- \\
    		\texttt{SubmitBid} &123350 &262933&159759 \\
    		\texttt{SetWinner} &82847&2872047&3487439 \\
    		\texttt{Withdraw} &20370&47112&--- \\
    		\texttt{Reset} & 402351&---&---\\
    		\bottomrule    
    	\end{tabular}
    \end{table}

    Compared to other sealed-bid auction constructions on top of Ethereum \cite{galal2018verifiable, galal2018succinctly}, Trustee achieves a significantly low and constant gas cost on the revelation of auction winner. The reason behind this is because most of the computations happen off-chain. Therefore, it costs the auctioneer less than 1 USD to deploy Trustee's smart contract $C$, start an auction, set the winner, and withdraw initial deposit. It has to be noted that, the initial deposit must be large enough to penalize malicious participants such as an auctioneer who corrupts $R$ to redirect inconsistent messages between $E$ and $C$, and a malicious winner who refuses to pay the second-highest price. Certainly, the value of the initial deposit should be proportional to the estimated value of the auctioned item.
    

    \section{Conclusion}
    \label{sec:conclusion}
     In this paper, we presented Trustee, an efficient and full privacy preserving Vickrey auction on top of Ethereum. In Trustee, we utilize Intel SGX to complement a smart contract in  Ethereum with confidential data processing, a desirable property they lack.  As a result, Trustee does not inherit the complexities of heavy cryptographic protocols such as ZKP and MPC. More precisely, Trustee fully preserves bids' privacy and maintains the auction winner correctness at a relatively cheap transaction fee. Furthermore, in Trustee, auctions take only two-rounds to finalize, where the first round is the provision of bids and the second one is the revelation of the winner. As a result, it is one round less than the (commit - reveal - prove) approach. Moreover, the major computations in Trustee happen on off-chain hosts, hence, it can be ported with minimum efforts to blockchains with inflexible scripting capabilities such as Bitcoin.

\bibliographystyle{plain}
\bibliography{biblio}

\begin{thebibliography}{10}

\bibitem{tokenmarket}
Digital assets in {Ethereum} blockchain.
\newblock \url{https://tokenmarket.net/blockchain/Ethereum/assets/}.

\bibitem{coinmarket}
Top 100 cryptocurrencies by market capitalization.
\newblock \url{https://coinmarketcap.com}, 2018.

\bibitem{al2018airtnt}
Mustafa Al-Bassam, Alberto Sonnino, Micha{\l} Kr{\'o}l, and Ioannis Psaras.
\newblock Airtnt: Fair exchange payment for outsourced secure enclave
  computations.
\newblock {\em arXiv preprint arXiv:1805.06411}, 2018.

\bibitem{anati2013innovative}
Ittai Anati, Shay Gueron, Simon Johnson, and Vincent Scarlata.
\newblock Innovative technology for {CPU} based attestation and sealing.
\newblock In {\em Proceedings of the 2nd international workshop on hardware and
  architectural support for security and privacy}, volume~13. ACM New York, NY,
  USA, 2013.

\bibitem{ben2014succinct}
Eli Ben-Sasson, Alessandro Chiesa, Eran Tromer, and Madars Virza.
\newblock Succinct non-interactive zero knowledge for a von neumann
  architecture.
\newblock In {\em USENIX Security Symposium}, pages 781--796, 2014.

\bibitem{benet2014ipfs}
Juan Benet.
\newblock Ipfs-content addressed, versioned, p2p file system.
\newblock {\em arXiv preprint arXiv:1407.3561}, 2014.

\bibitem{bentov2017tesseract}
Iddo Bentov, Yan Ji, Fan Zhang, Yunqi Li, Xueyuan Zhao, Lorenz Breidenbach,
  Philip Daian, and Ari Juels.
\newblock Tesseract: Real-time cryptocurrency exchange using trusted hardware.
\newblock {\em IACR Cryptology ePrint Archive}, 2017:1153, 2017.

\bibitem{bertoni2013keccak}
Guido Bertoni, Joan Daemen, Micha{\"e}l Peeters, and Gilles Van~Assche.
\newblock Keccak.
\newblock In {\em Annual international conference on the theory and
  applications of cryptographic techniques}, pages 313--314. Springer, 2013.

\bibitem{blass2018strain}
Erik-Oliver Blass and Florian Kerschbaum.
\newblock Strain: A secure auction for blockchains.
\newblock In {\em European Symposium on Research in Computer Security}, pages
  87--110. Springer, 2018.

\bibitem{bos2014elliptic}
Joppe~W Bos, J~Alex Halderman, Nadia Heninger, Jonathan Moore, Michael Naehrig,
  and Eric Wustrow.
\newblock Elliptic curve cryptography in practice.
\newblock In {\em International Conference on Financial Cryptography and Data
  Security}, pages 157--175. Springer, 2014.

\bibitem{brandenburger2018blockchain}
Marcus Brandenburger, Christian Cachin, R{\"u}diger Kapitza, and Alessandro
  Sorniotti.
\newblock Blockchain and trusted computing: Problems, pitfalls, and a solution
  for {Hyperledger Fabric}.
\newblock {\em arXiv preprint arXiv:1805.08541}, 2018.

\bibitem{secp256k1}
Daniel R.~L. Brown.
\newblock Standards for efficient cryptography sec 2: Recommended elliptic
  curve domain parameters.
\newblock \url{http://www.secg.org/sec2-v2.pdf}, 2010.

\bibitem{chen2018sgxpectre}
Guoxing Chen, Sanchuan Chen, Yuan Xiao, Yinqian Zhang, Zhiqiang Lin, and Ten~H
  Lai.
\newblock {SGXPECTRE}attacks: Leaking enclave secrets via speculative
  execution.
\newblock {\em arXiv preprint arXiv:1802.09085}, 2018.

\bibitem{cheng2018ekiden}
Raymond Cheng, Fan Zhang, Jernej Kos, Warren He, Nicholas Hynes, Noah Johnson,
  Ari Juels, Andrew Miller, and Dawn Song.
\newblock Ekiden: A platform for confidentiality-preserving, trustworthy, and
  performant smart contract execution.
\newblock {\em arXiv preprint arXiv:1804.05141}, 2018.

\bibitem{galal2018succinctly}
Hisham~S Galal and Amr~M Youssef.
\newblock Succinctly verifiable sealed-bid auction smart contract.
\newblock In {\em Data Privacy Management, Cryptocurrencies and Blockchain
  Technology}, pages 3--19. Springer, 2018.

\bibitem{galal2018verifiable}
Hisham~S Galal and Amr~M Youssef.
\newblock Verifiable sealed-bid auction on the {Ethereum} blockchain.
\newblock In {\em International Conference on Financial Cryptography and Data
  Security}, pages 265--278. Springer, 2018.

\bibitem{gayoso2010survey}
V{\'\i}ctor Gayoso~Mart{\'\i}nez, Luis Hern{\'a}ndez~Encinas, and Carmen
  S{\'a}nchez~{\'A}vila.
\newblock A survey of the elliptic curve integrated encryption scheme.
\newblock 2010.

\bibitem{gruss2017strong}
Daniel Gruss, Julian Lettner, Felix Schuster, Olya Ohrimenko, Istvan Haller,
  and Manuel Costa.
\newblock Strong and efficient cache side-channel protection using hardware
  transactional memory.
\newblock In {\em USENIX Security Symposium}, pages 217--233, 2017.

\bibitem{lee2017inferring}
Sangho Lee, Ming-Wei Shih, Prasun Gera, Taesoo Kim, Hyesoon Kim, and Marcus
  Peinado.
\newblock Inferring fine-grained control flow inside {SGX} enclaves with branch
  shadowing.
\newblock In {\em 26th USENIX Security Symposium, USENIX Security}, pages
  16--18, 2017.

\bibitem{lind2016teechan}
Joshua Lind, Ittay Eyal, Peter Pietzuch, and Emin~G{\"u}n Sirer.
\newblock Teechan: Payment channels using trusted execution environments.
\newblock {\em arXiv preprint arXiv:1612.07766}, 2016.

\bibitem{matetic2017rote}
Sinisa Matetic, Mansoor Ahmed, Kari Kostiainen, Aritra Dhar, David Sommer,
  Arthur Gervais, Ari Juels, and Srdjan Capkun.
\newblock Rote: Rollback protection for trusted execution.
\newblock {\em IACR Cryptology ePrint Archive}, 2017:48, 2017.

\bibitem{milutinovic2016proof}
Mitar Milutinovic, Warren He, Howard Wu, and Maxinder Kanwal.
\newblock Proof of luck: An efficient blockchain consensus protocol.
\newblock In {\em Proceedings of the 1st Workshop on System Software for
  Trusted Execution}, page~2. ACM, 2016.

\bibitem{schwarz2017malware}
Michael Schwarz, Samuel Weiser, Daniel Gruss, Cl{\'e}mentine Maurice, and
  Stefan Mangard.
\newblock Malware guard extension: Using {SGX} to conceal cache attacks.
\newblock In {\em International Conference on Detection of Intrusions and
  Malware, and Vulnerability Assessment}, pages 3--24. Springer, 2017.

\bibitem{seo2017sgx}
Jaebaek Seo, Byoungyoung Lee, Seong~Min Kim, Ming-Wei Shih, Insik Shin, Dongsu
  Han, and Taesoo Kim.
\newblock {SGX}-shield: Enabling address space layout randomization for sgx
  programs.
\newblock In {\em NDSS}, 2017.

\bibitem{shih2017t}
Ming-Wei Shih, Sangho Lee, Taesoo Kim, and Marcus Peinado.
\newblock T-{SGX}: Eradicating controlled-channel attacks against enclave
  programs.
\newblock In {\em Proceedings of the 2017 Annual Network and Distributed System
  Security Symposium (NDSS), San Diego, CA}, 2017.

\bibitem{shinde2016preventing}
Shweta Shinde, Zheng~Leong Chua, Viswesh Narayanan, and Prateek Saxena.
\newblock Preventing page faults from telling your secrets.
\newblock In {\em Proceedings of the 11th ACM on Asia Conference on Computer
  and Communications Security}, pages 317--328. ACM, 2016.

\bibitem{tran2017obscuro}
Muoi Tran, Loi Luu, Min~Suk Kang, Iddo Bentov, and Prateek Saxena.
\newblock Obscuro: A bitcoin mixer using trusted execution environments.
\newblock {\em IACR Cryptology ePrint Archive}, 2017:974, 2017.

\bibitem{vanbulck2018foreshadow}
Jo~Van~Bulck, Marina Minkin, Ofir Weisse, Daniel Genkin, Baris Kasikci, Frank
  Piessens, Mark Silberstein, Thomas~F. Wenisch, Yuval Yarom, and Raoul
  Strackx.
\newblock Foreshadow: Extracting the keys to the {Intel SGX} kingdom with
  transient out-of-order execution.
\newblock In {\em Proceedings of the 27th {USENIX} Security Symposium}.
  {USENIX} Association, August 2018.

\bibitem{weisse2018foreshadowNG}
Ofir Weisse, Jo~Van~Bulck, Marina Minkin, Daniel Genkin, Baris Kasikci, Frank
  Piessens, Mark Silberstein, Raoul Strackx, Thomas~F. Wenisch, and Yuval
  Yarom.
\newblock {Foreshadow-NG}: Breaking the virtual memory abstraction with
  transient out-of-order execution.
\newblock {\em Technical report}, 2018.

\bibitem{wood2014ethereum}
Gavin Wood.
\newblock Ethereum: A secure decentralised generalised transaction ledger.
\newblock {\em Ethereum Project Yellow Paper}, 151:1--32, 2014.

\bibitem{xu2015controlled}
Yuanzhong Xu, Weidong Cui, and Marcus Peinado.
\newblock Controlled-channel attacks: Deterministic side channels for untrusted
  operating systems.
\newblock In {\em Security and Privacy (SP), 2015 IEEE Symposium on}, pages
  640--656. IEEE, 2015.

\bibitem{mbedtls}
Fan Zhang.
\newblock mbedtls-sgx: a {TLS} stack in {SGX}.
\newblock \url{https://github.com/bl4ck5un/mbedtls-SGX}, 2016.

\bibitem{zhang2016town}
Fan Zhang, Ethan Cecchetti, Kyle Croman, Ari Juels, and Elaine Shi.
\newblock Town crier: An authenticated data feed for smart contracts.
\newblock In {\em Proceedings of the 2016 ACM SIGSAC conference on computer and
  communications security}, pages 270--282. ACM, 2016.

\end{thebibliography}
\end{document}